\documentclass[twocolumn,natbib]{svjour3}          % twocolumn; this is how it will look like
\usepackage{graphicx}
\usepackage{mathptmx}      % use Times fonts if available on your TeX system
\usepackage[utf8x]{inputenc}
\usepackage{amsmath}
\usepackage{amsfonts}
\usepackage{amssymb}
\usepackage{siunitx}
\usepackage{placeins}
\journalname{Microfluidics and Nanofluidics}
\begin{document}

\title{Self-assembly of coated microdroplets at the sudden expansion of a microchannel\thanks{This work was supported by a Daphne Jackson Fellowship funded by the The University of Manchester and the Engineering and Physical Sciences Research Council (EPSRC).}
}

\titlerunning{Coated microdroplet clustering}        % if too long for running head

\author{Kerstin Schirrmann \and Gabriel C\'aceres-Aravena \and Anne Juel
}

%\authorrunning{Schirrmann \and Caceres \and Juel} % if too long for running head

\institute{K. Schirrmann \and G. C\'arceres-Aravena \and A. Juel \at
              The University of Manchester, Department of Physics \& Astronomy, Manchester, M13 9PL, UK\\
              %Tel.: +44 161 2754071\\
              \email{kerstin.schirrmann@manchester.ac.uk,\\ anne.juel@manchester.ac.uk}           %  \\
}

\date{Received: date / Accepted: date}

\maketitle

\begin{abstract}

We report observations of the self-assembly of coated droplets into regular clusters at the sudden expansion of a microfluidic channel. A double emulsion consisting of a regular train of coated microdroplets was created upstream of the channel expansion, so that the inter-drop distance, droplet length, velocity and coating thickness could be varied by imposing different inlet pressures, albeit not independently. Provided that the enlarged channel remains sufficiently confined to prohibit propagation in double file, droplets can assemble sequentially into regular linear clusters at the expansion. Droplets join a cluster via the coalescence of their coating film with that of the group ahead. This coalescence occurs when the droplets approach each other to within a critical distance at the expansion, enabled by hydrodynamic interactions within the train. Clusters comprising a finite number of droplets are obtained because reconfiguration of the droplet assembly during coalescence increases the distance to the following droplet. Decreasing the inter-drop distance increases the cluster size up to a maximum value beyond which continuous clusters form. Formalising these observations in a simple model reveals that clusters of any size are possible but that they occur for increasingly narrow ranges of parameter values. Our experimental observations suggests that background experimental fluctuations limit the maximum discrete cluster size in practice. This method of self-assembly offers a robust alternative to flow focusing for encapsulating multiple cores in a single coating film and the potential to build more complex colloidal building blocks by de-confining the clusters. 

\end{abstract}

\section{Introduction}
\label{sec:introduction}

Droplet microfluidics rely on the controlled generation and manipulation of microdroplets with a high degree of reproducibility~\citep{Anna2016} in order to provide a platform for high-throughput applications in material science, chemistry and biotechnology. Individual droplets encapsulate reagents and/or biological material for protein engineering,  enzyme screening or DNA/RNA sequencing testing~\citep{Pompano2011,Seemann2012,Vladisavljevic2013,Convery2019,Payne2020}. However, the complex dynamics arising from coupling multiple moving droplets with channel confinement continues to reveal new fundamental phenomenology~\citep{Tabeling2014}. In this paper, we report experimental observations of the self-assembly of microdroplets downstream of a sudden expansion in a microfluidic channel into a variety of structures.

Trains of microdroplets are typically generated by breaking up continuous streams of liquids to form an emulsion, which can in turn be manipulated by a variety of passive or active microfluidic strategies~\citep{Seemann2012}. 
Passive manipulation relies on features of the microchannel geometry to drive flow-induced actuation of droplets. Whereas undesirable geometrical defects in the microfluidic channel can disturb a uniform train of droplets and induce coalescence, controlled geometrical features can be used to manipulate the droplet train. A localised obstruction in a channel may lead to either coalescence of two droplets or break-up of a droplet into smaller parts depending on its positioning and size~\citep{Vladisavljevic2013}. The merging of two droplets can be achieved by smoothly expanding or contracting the microfluidic channel~\citep{Bremond2008,Seemann2012,Tabeling2014} and multiple forms of coalescence of two droplets downstream of a sudden channel expansion have been characterised experimentally by \citet{Shen2017}. \citet{Bremond2008} find that two droplets tend to coalesce upon separation occuring after collision. The coalescence mechanism, which overcomes the stabilising effect of surfactants, can induce cascades of coalescence in closely packed droplets, a phenomenon also observed by \citet{Jose2012} in a different setting. \citet{Tan2007} showed that a uniform train of droplets at a trifurcating junction comprising an expanded chamber could redistribute and fuse consistently to yield coalesced groups of up to six droplets, by regulating the downstream flow rates. Multiple coalescence scenarios were explored which were triggered by the deformation of the front droplet re-entering a more confined channel past the expansion chamber. 

Passive actuation can also result in the self-assembly of multiple individual droplets or bubbles. A striking example is provided by the microfluidic generation of foams and emulsions into gliding, ordered two-dimensional arrays by coupling flow and geometric confinement~\citep{Seo2005,Marmottant2009}. Trains of droplets in quasi-two-dimensional microchannels can exhibit dynamics similar to one-dimensional crystal-like structures such as the emergence of collective normal vibrational modes~\citep{Beatus2006, Beatus2012}. This is because the motion of the droplets induces dipolar flow fields, which mediate hydrodynamic interaction between the droplets. Self-assembly of droplets into arrays for the rapid analysis of simultaneous reactions can also be achieved by providing anchoring sites for each droplet in the form of a notch in the channel~\citep{Pompano2011}. Recently, \citet{Shen2016} exploited hydrodynamic interactions and
adhesive depletion forces (due to surfactant above the critical micelle concentration) to assemble colloidal building blocks of varied shapes. This was achieved by breaking up flowing plugs of liquid into closely packed droplets at the sudden expansion of a microchannel into a wider and deeper chamber. 

In this paper, we focus on coated microdroplets which consist of an inner droplet encapsulated by a liquid coating. Trains of coated droplets are typically produced with two successive droplet generation devices~\citep{Hennequin2009}, which can provide a vast range of dynamics~\citep{Anna2016,Wang2020}. The coated droplets may be cured into micro-capsules if the coating phase can be polymerised into an elastic membrane, or combine reagents for chemical reactions~\citep{Utada2005}. We explore the sequential self-assembly of coated microdroplets into clusters at a sudden expansion in a microfluidic channel (see figure \ref{fig:setup}). Clusters of droplets are formed through the coalescence of their coating films, which are either one-dimensional arrays or three-dimensional structures depending on the level of confinement imposed by the microchannel. We explore the role of the droplet-train properties upstream of the channel expansion and the droplet-velocity reduction at the expansion in determining the number of droplets encapsulated in each cluster. We vary the size of the channel expansion ratio and find increasing finite-cluster sizes up to the limit where the downstream channel can accommodate droplets in double files. 

The fluid dynamics of droplet trains in confined channels is challenging to model. A long bubble in a liquid-filled channel of rectangular cross-section acts like a leaky piston where the motion of the bubble is regulated by the large drag exterted by the sidewalls of the channel so that liquid bypasses the bubble by flowing through the channel corners~\citep{Wong1995b}. The nature of the recirculation flows ahead of long bubbles determines their speed relative to the mean flow depending on imposed flow rate and channel aspect ratio~\citep{deLozar2008}. Considering droplets of modest length brings additional complexity and experiments have shown that individual droplets can move either faster or slower than the mean flow in confined rectangular geometries~\citep{Jakiela2011, Sessoms2009}. Numerical simulations~\citep{Wang2011} have identified the droplet length, capillary number and viscosity ratio between phases as important parameters in a square capillary. For trains of droplets, the hydrodynamic interaction between the droplets further influences the droplet velocity with vortex pairs forming both inside and between the droplets~\citep{Sarrazin2008,Labrot2009,Jakiela2016}. We are not aware of similar studies on coated droplet trains, where different flows patterns will be generated within the coated microdroplet compared with simple droplets. We will therefore rely on experimental measurements of the droplet velocity and length both upstream and downstream of our expansion in order to interpret the organised clustering we observe.  

The paper is organised as follows. We describe our microfluidic device and experimental methods in section~\ref{sec:methods}. Results are presented in section~\ref{sec:results}, where we analyse our experimental observations of the self-assembly into droplet clusters in section~\ref{sec:formation} and formalise our experimental observations with a simple model in section~\ref{sec:model}. Conclusions are given in section~\ref{sec:conclusion}.

\section{Experimental methods} \label{sec:methods}
\subsection{Microfluidic device}
\label{sec:device}

\begin{figure}[htb]
	\centering
	\includegraphics[width=\columnwidth]{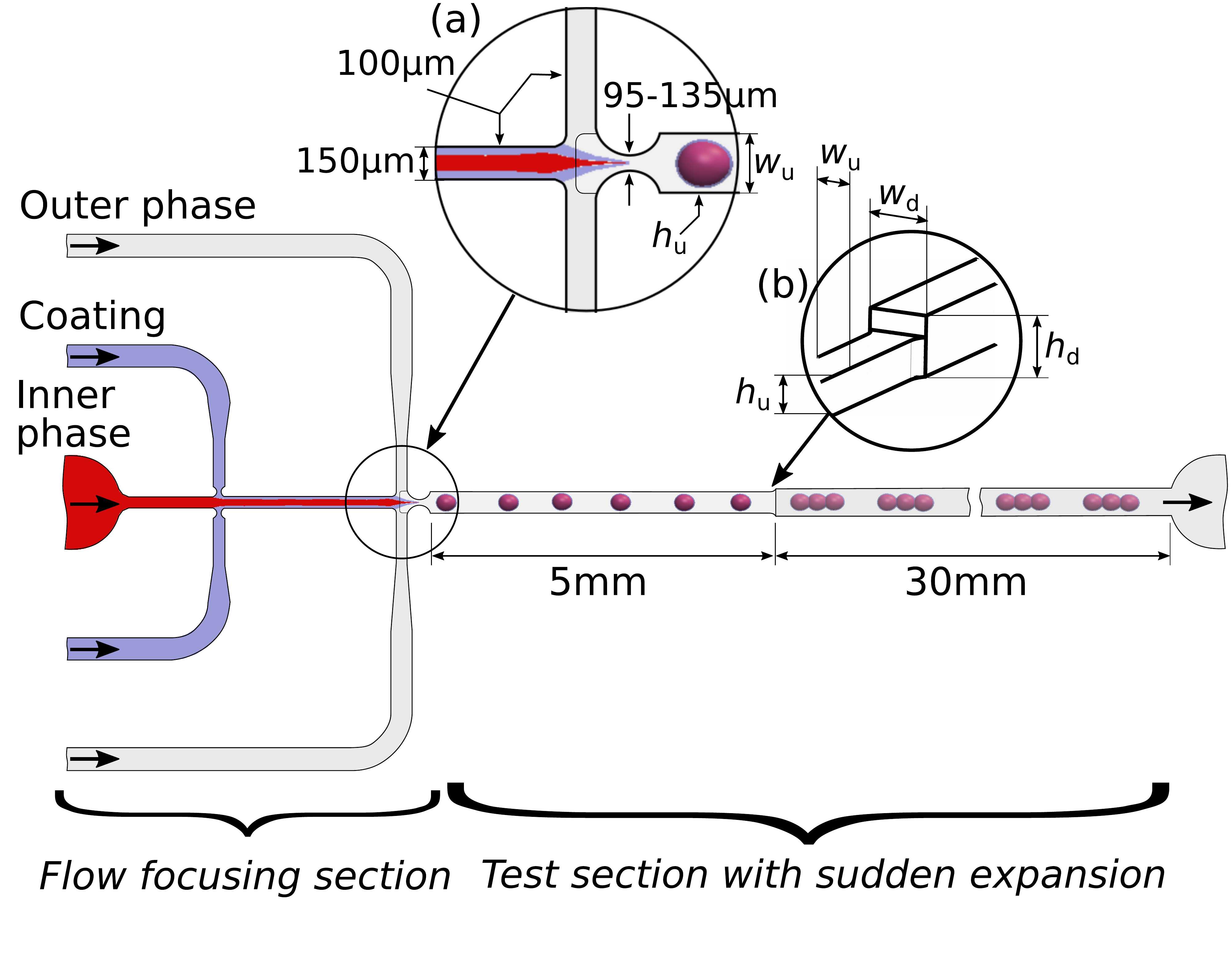}
	\caption{{Schematic diagram of the microfluidic device. The coating phase (blue, perfluorinated polyether with fluorosurfactant) is injected at the first cross-junction so that it encapsulates the inner phase (red, dyed aqueous glycerol solution) flowing along the straight channel leading to a core-annular flow. The outer phase (grey, silicone oil) is injected at the second cross-junction which is adjacent to the flow-focusing constriction (inset~(a) where coated microdroplets are pinched off to form a regular train suspended in the outer phase, which propagates through the test section. The coated microdroplets forming the train may cluster depending on the flow parameters as they propagate past the sudden channel expansion (inset~(b)). }}\label{fig:setup}
\end{figure}

%description of device and fluids
A schematic diagram of the microfluidic device is shown in figure~\ref{fig:setup}. It comprises two elements in series: a flow-focusing section to generate regular trains of coated microdroplets followed by a test section which consists of a straight channel of rectangular cross-section with a sudden expansion (inset (b)) located \num{5}~mm downstream of the flow focusing constriction (inset (a)). The channel downstream of the expansion is \SI{30}{\milli\meter} long. We used four microfluidic devices with similar flow focusing elements and a rectangular channel of height $h_{\mathrm{u}} = \SI{260}{\micro\meter}$ and width $w_{\mathrm{u}}=\SI{310}{\micro\meter}$ or $w_\mathrm{u}=\SI{330}{\micro\meter}$ which was expanded into channels of square cross-section with $h_{\mathrm{d}} =w_{\mathrm{d}} =  345,\; 370,\; 415\; \mathrm{and}\;\SI{490}{\micro\meter}$, respectively, each uniform to within $\pm \SI{3}{\micro\meter}$.
We define the expansion factor $\alpha$ as the ratio of the cross-sections so that the microfluidic devices had expansions with $\alpha = h_{\mathrm{d}}\,w_{\mathrm{d}}/(h_\mathrm{u}\, w_\mathrm{u})= 1.5, \,1.7,\; 2.0\; \mathrm{and}\; 3.0$, respectively.

The flow focusing section shown in figure~\ref{fig:setup} is similar to that described by \citet{Hennequin2009}. The coating phase (blue) was added through a cross-junction to the inner phase (red) flowing along the main straight channel, so that the inner phase was encapsulated by the coating phase in a core-annular flow. A second cross-junction immediately followed by a flow-focusing constriction (inset (a)) enabled the injected outer phase (grey) to break the core annular stream into coated microdroplets of the inner phase suspended in the outer phase. The generation of regular trains of coated microdroplets (or double emulsion) using this sequence relied on favourable wetting properties between the fluids phases and between the fluids and the channel boundaries. The inner phase (subscript i) was an aqueous solution of glycerol (Sigma Aldrich, 57\% by volume, density $\rho_\mathrm{i} =\SI{1.16}{\gram\per\milli\litre}$ at \SI{25}{\celsius}) dyed with carminic acid. The coating fluid (subscript~c) was a perfluorinated polyether (\texttrademark{Galden} HT135, Solvay, density $\rho_\mathrm{c} =\SI{1.72}{\gram\per\milli\litre}$ at \SI{25}{\celsius}) with a 2\% concentration by weight of 008 Fluorosurfactant (RAN Biotechnologies), which was above the critical micelle concentration of $<\SI{0.1}{\percent}$ by weight. The outer phase (subscript~o) which acted as the suspending fluid was silicone oil (Sigma Aldrich, viscosity $\nu_\mathrm{o}=\SI{20}{\centi St}$, density $\rho_\mathrm{o} =\SI{0.95}{\gram\per\milli\litre}$ at \SI{25}{\celsius}).

% fabrication
The microfluidic devices were made from PDMS (Polydimethylsiloxane, Sylgard 184, Dow Corning), which was degassed before moulding. Micromilled brass negatives (Datron M6) were used to mould the device geometry shown in figure \ref{fig:setup}. After curing, access holes were punched into the resulting open channels. These were then bonded onto glass slides with a cured PDMS coating. For bonding, both parts were oxidised using oxygen plasma treatment (Henniker plasma HPT-100) at \SI{50}{\percent} power for \SI{20}{\second} and immediately brought into contact. To restore lipophilic wetting behaviour after the bonding process, the devices were aged for at least 48 hours at \SI{50}{\celsius} and primed for at least \SI{30}{\minute} with silicone oil.

% drive (pressure controller) 
The flow was driven by pressurising the three inlets of the microfluidic device (inner phase, coating, outer phase) relative to the outlet of the device, which was left open at atmospheric pressure. The pressure at each inlet was controlled independently by connecting the inlet to a fluid container of constant pressure set by a pressure controller (Elveflow Mk3+ \num{0}-\SI{2}{\bar}, Elvesys). Flow devices with a resistance much larger than that of the flow focusing device were introduced between the pressurised reservoirs and the inlets of the microfluidic device in order to maintain a constant inlet flow rate by decoupling the inlet from the dynamics of the microfluidic device. These flow resistors consisted of PDMS microchannels of square cross-section with \SI{100(10)}{\micro\meter} width and height and lengths of \SI{57}{\milli\meter} for the inner and outer phase inlets, and \SI{270}{\milli\meter} for the coating inlet.

\subsection{Visualisation and image analysis}
\label{sec:visualisation}

\begin{figure}[!ht]
	\centering
	\includegraphics[width=\columnwidth]{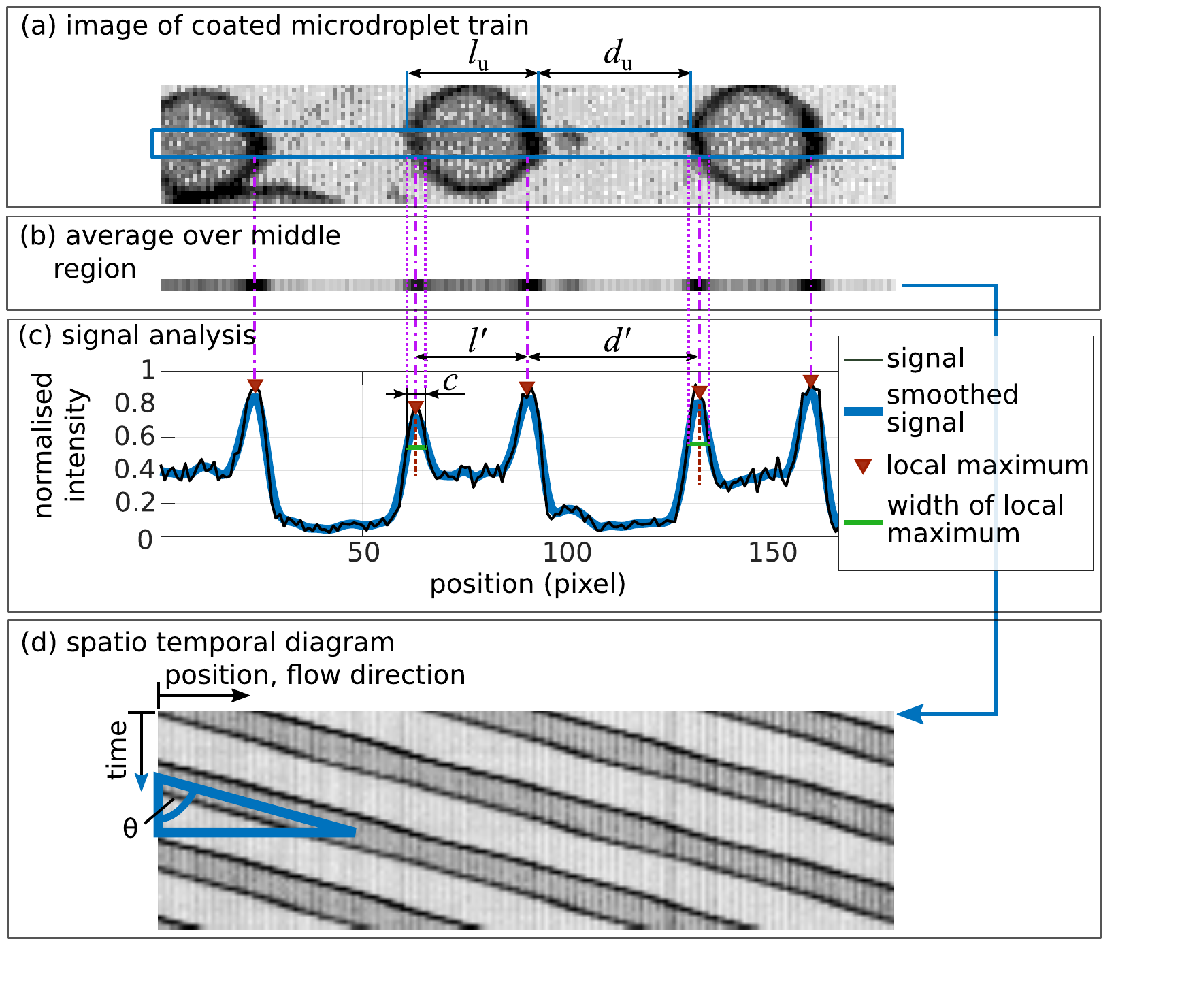}
	\caption{{Analysis of top-view images of a coated micropdroplet train. (a) Close-up image of a micro-droplet train. (b) Vertical average of the region enclosed by the blue box in image (a). (c) Normalised pixel intensity as a function of streamwise position from image (b). The distance between local maxima correspond to approximate measures of the droplet length $l'$ and inter-droplet distance $d'$ in alternation. We define the coating thickness parameter $c$ as the width of the peaks at half height. The corrected metrics $l_u=l'+c$ and $d_u=d'-c$ correspond to the total length of the droplet including coating and the actual inter-droplet distance. (d) Spatio-temporal diagram obtained by vertical concatenation of successive averaged images similar to that shown in (b). The regularly spaced, linear droplet paths indicate that the spacing between the microdroplets in the train is approximately constant and that the microdroplets propagate with constant speed. }} \label{fig:signals}
\end{figure}

%recording
The coated microdroplet trains propagating in the microfluidic devices were imaged in top-view with a monochrome CMOS camera (PCO 1200hs, 10x macro lens) at 100 frames per second with a resolution of \SI{103}{px\per\mm}. The channels were backlit using a custom-made LED light panel. A typical train of droplets is shown in figure~\ref{fig:signals}(a), where the dyed aqueous inner phase of the droplet is surrounded by a darker coating layer and the outer phase is bright. 

%analysis
We characterised the trains of coated microdroplets upstream of the expansion by the droplet length $l_\mathrm{u}$, the distance between droplets $d_\mathrm{u}$, a coating thickness parameter $c$ and the droplet speed $v_\mathrm{u}$ using image processing routines programmed in Matlab. Each frame was cropped to retain a 7 pixel-wide region (\SI{68}{\micro\meter}) indicated by the blue box in figure~\ref{fig:signals}(a) and averaged vertically to yield the line in figure~\ref{fig:signals}(b). The grey values in this line were inverted, normalised, and smoothed with a Gaussian weighted-moving-average filter to generate the intensity plot shown in figure~\ref{fig:signals}(c). The peaks in the signal indicate the coating, and the inner phase is associated with higher intensity values than the suspending outer phase. The distance between local maxima of the signal correspond in alternation to approximate measures of the droplet length $l'$ and the distance between droplets $d'$. We define the coating thickness parameter $c$ as the width of each peak at half height. Hence, the total droplet length is $l_u=l'+c$ and the actual inter-drop distance is $d_u =d'-c$; see figure~\ref{fig:signals}(a,c). This method of analysis gave consistent results for all experiments, including those where the inter-droplet distance was very small or where the image contrast was reduced. 

Figure~\ref{fig:signals}d shows a spatio-temporal plot with position along the horizontal axis and time down the vertical axis. To obtain this plot, we vertically concatenated the averaged centre regions of successive images (see figure~\ref{fig:signals}(b)). In the spatio-temporal diagram, the droplet paths appear as dark bands with darker edges marking the coating. The regularly spaced droplet paths of approximately constant bandwidth indicate a regular droplet train. The droplet paths are approximately linear and oriented at an angle $\theta$ with respect to the vertical axis which is measured to yield the approximately constant droplet speed $v_u=\tan \theta$. We refer to figure~\ref{fig:2process+xts} for spatio-temporal diagrams that include the expansion. In these, the width of the bands downstream of the expansion was used to calculate the cluster size and their angle with respect to the vertical gave the cluster speed. 

\subsection{Parameter variations of the droplet train upstream of the expansion}
\label{sec:parameter}

We varied the three inlet pressures to obtain droplet trains with different droplet size, inter-droplet distance, coating thickness and velocity. The relation between the inlet pressures and the droplet train properties was set by complex dynamics in the flow focusing section and was not studied systematically. This meant that we were unable to vary a single droplet-train parameter while keeping the other three constant.
We obtained regular trains of coated droplets only for a limited range of pressure combinations (overall ranges: outer phase \num{320} to \SI{1480}{\milli\bar}, coating phase \num{33} to \SI{200}{\milli\bar}, inner phase \num{50} to \SI{510}{\milli\bar}) and avoided pressure values leading to other three-phase flows, e.g. droplets of the inner and coating phases in alternation, droplets of the coating phase amongst coated droplets or multiple inner-phase droplets encapsulated in a coating droplet as recently explored numerically by \citet{Wang2020}.

Table~\ref{tab:exp_values} gives an overview of the upstream droplet train parameters that were explored within the pressure range for regular trains of coated droplets and expansion factors $\alpha \le 2.0$. The inter-droplet distance $d_u$ could be varied most systematically through changes in the pressure difference between the outer and inner phases, although the other parameters $l_u$, $c$ and $v_u$ would also vary but to a lesser extent. The accessible range of droplet length differed between channel geometries due to small discrepancies in the flow focusing constriction. Variations in the droplet length were achieved by increasing/decreasing all pressures, which also strongly influenced the velocity. Variations in the value of the coating thickness parameter was naturally limited by the requirement to generate trains of individual coated droplets. For each experiment performed with fixed pressure values, all droplet train parameters remained approximately constant, with typical standard deviations of \SI{8}{\micro\meter} for droplet length and \SI{12}{\micro\meter} for inter-droplet distance, which was typically on the order of a few percents (see table~\ref{tab:exp_values}). 

Table~\ref{tab:exp_values} also shows typical values of the upstream capillary number $Ca=\nu_{\mathrm o} \rho_{\mathrm o} v_{\mathrm u}/\sigma_{\mathrm oc}$ -- a measure of the ratio of viscous to surface tension force -- which are on the order of $10^{-3}$, where $\sigma_{\mathrm{oc}}=\SI{36}{\milli\newton\per\meter}$ is the interfacial tension between the coating and outer phase.  The Bond number $Bo=\Delta \rho g R^2/\sigma_{\rm oc}$ -- a measure of the ratio of gravitational to surface tension forces -- takes values $Bo_{\mathrm{max}} < \num{6.3e-3}$ for a spherical droplet of diameter $2R=\SI{0.415}{\milli\meter}$ including a \SI{52}{\micro\meter} coating film (corresponding to $w_{\mathrm{d}}$ for $\alpha=2$) indicating that gravitational forces are insignificant.
The droplet Reynolds number $Re=v_{\rm u} h_{\rm u} / \nu_o$ -- a measure of the ratio of inertial to viscous forces -- takes values $0.02 \le Re \le 0.1$, which means that inertial effects are broadly negligible, with the possible exception of the largest droplet velocities.

\begin{table}
\caption{{Range of properties accessible experimentally for the train of coated microdroplets upstream of the expansion as function of the expansion factor.}}\label{tab:exp_values}
\begin{tabular}{l c c c }
\hline\noalign{\smallskip}
expansion factor $\alpha$&1.5&1.7&2.0\\ 
\noalign{\smallskip}\hline\noalign{\smallskip}
$d_{\mathrm{u}}$ (\si{\micro\meter})&\numrange{29}{305}&\numrange{62}{588}&\numrange{165}{814}\\ 
$l_{\mathrm{u}}$ (\si{\micro\meter})&\numrange{364}{573}&\numrange{285}{450}&\numrange{282}{370}\\ 
$c$ (\si{\micro\meter})&\numrange{45}{87}&\numrange{53}{78}&\numrange{47}{56}\\ %\hline
$v_{\mathrm{u}}$ (\si{\milli\meter\per\second})&\numrange{2.8}{8.5}&\numrange{1.9}{7.1}&\numrange{3.4}{4.0}\\ 
\noalign{\smallskip}\hline\noalign{\smallskip}
$Ca$ ($\nu_{\mathrm o} \rho_{\mathrm o} v_{\mathrm{u}}/\sigma_{\mathrm{oc}}$)&\numrange{0.002}{0.005}&\numrange{0.001}{0.004}&\num{0.002}\\ %\hline
$Re$ ($v_{\mathrm{u}}h_{\mathrm{u}}/\nu_{\mathrm{o}}$)&\numrange{0.04}{0.1}&\numrange{0.02}{0.09}&\numrange{0.04}{0.05}\\ 
\noalign{\smallskip}\hline
\end{tabular}
\end{table}

\section{Results}
\label{sec:results}

\subsection{Cluster formation}
\label{sec:formation}

\begin{figure}
	\centering
	\includegraphics[width=\columnwidth]{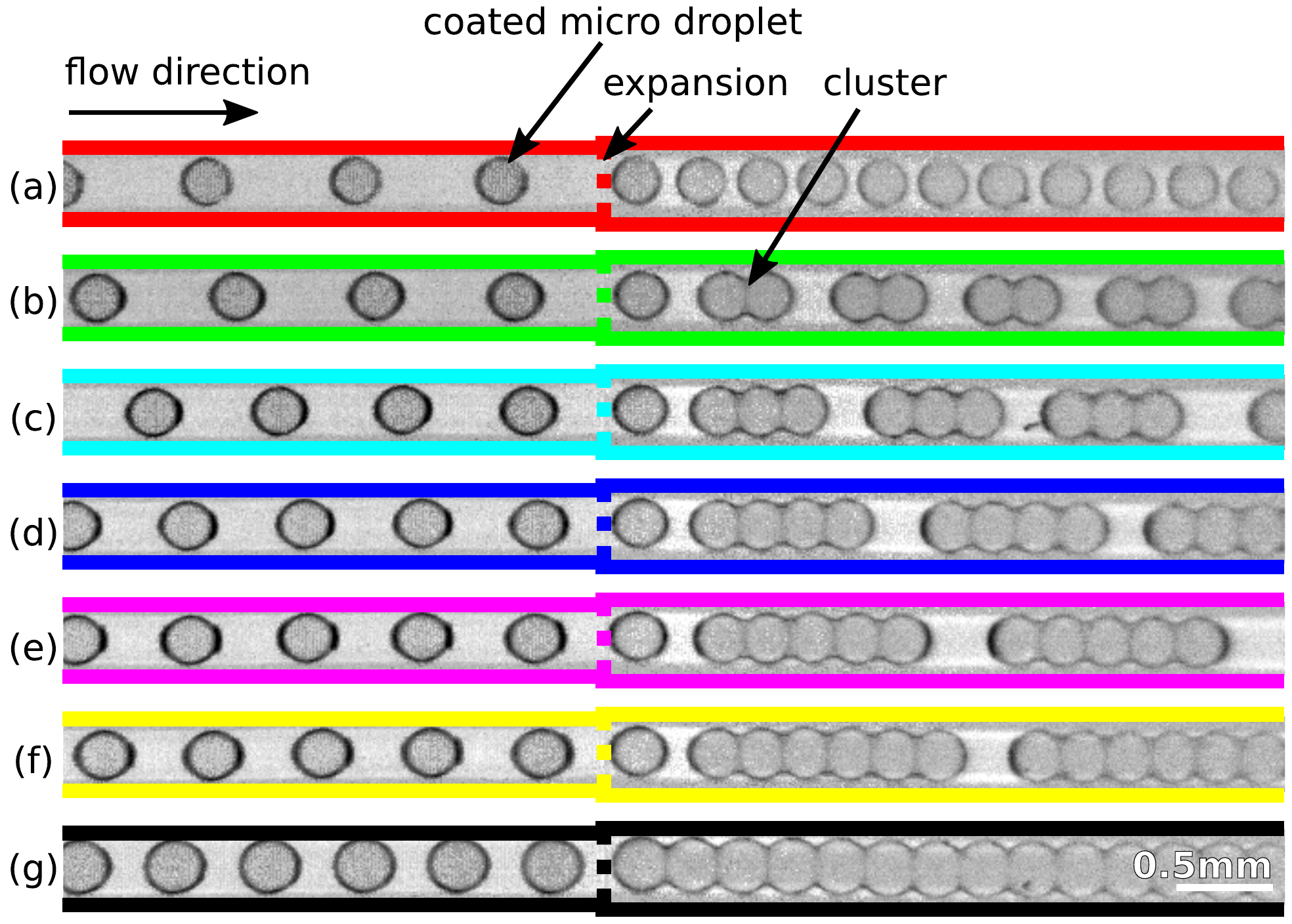}
	\caption{{Series of snapshots illustrating the range of behaviour of trains of coated microdroplets flowing through a microchannel with a sudden expansion of factor $\alpha=2.0$, indicated by a vertical dotted line.The single-droplet train upstream of the expansion is regular but the train parameters -- inter-drop distance, droplet length, coating film thickness and droplet velocity -- vary between images. The most significant variation is in the inter-drop distance which decreases monotonically from (a) to (g). In (a), a single-droplet train is retained downstream of the expansion, while in (b -- f) the droplets have assembled into constant-size clusters of between two to six droplets through the coalescence of their coating films. By contrast, a long, continuous cluster is formed in (g). Background subtraction and contrast adjustment have been applied to all images. Coloured lines outline the channel geometry, with each colour indicating a different cluster size.}} \label{fig:gall}
\end{figure}

The sudden expansion of the microchannel (see figure \ref{fig:setup}) provides a passive means for a small number of droplets to self-assemble sequentially into clusters through the coalescence of their coating films. Figure~\ref{fig:gall} shows a series of snapshots centred on the expansion (with $\alpha =2.0$) which illustrate the range of behaviour observed experimentally (video in online supplement). Upstream of the expansion, the droplets propagate in a regular train of individual droplets. The four droplet train parameters  upstream of the expansion introduced in section~\ref{sec:visualisation} 
%--  inter-droplet distance, droplet length, coating film thickness and droplet velocity -- 
vary between images but the most significant parameter variation is in the inter-drop distance, which decreases monotonically from (a) to (g). In (a), the inter-drop distance is sufficiently large so that a train of regularly-spaced individual droplets is retained downstream of the expansion. As each droplet moves through the expansion into the enlarged channel, its velocity decreases mainly because of the reduction in the mean velocity of the suspending fluid. However, in (a) this reduction is not sufficient to enable the next droplet in the train to catch up with the droplet ahead while still propagating with a larger velocity. 
In figure~\ref{fig:gall}(b--f), the droplets reorganise downstream of the expansion into regularly-spaced, steadily propagating clusters of constant length. Each cluster contains individual droplets of the inner phase encapsulated in a shared coating film and separated by a thin film of coating fluid. The number of droplets in each cluster appears to increase with decreasing inter-drop distance from two to six droplets. We did not observe the formation of regular trains of finite-size clusters containing more than six droplets. Instead, for sufficiently small inter-drop distance, the droplets typically coalesced into a continuous cluster at the expansion, as shown in figure \ref{fig:gall}(g). We also observed trains of clusters of varying size near the threshold in inter-drop distance separating two cluster sizes. We found qualitatively similar clustering behaviour to that shown in figure \ref{fig:gall} in microchannels with smaller expansion factors, $\alpha=1.5$ and $1.7$. Increasing the expansion factor to $\alpha=3$ led to a wider range of droplet dynamics that will be discussed briefly at the end of this section.

\begin{figure*}
	\includegraphics[width=1.0\textwidth]{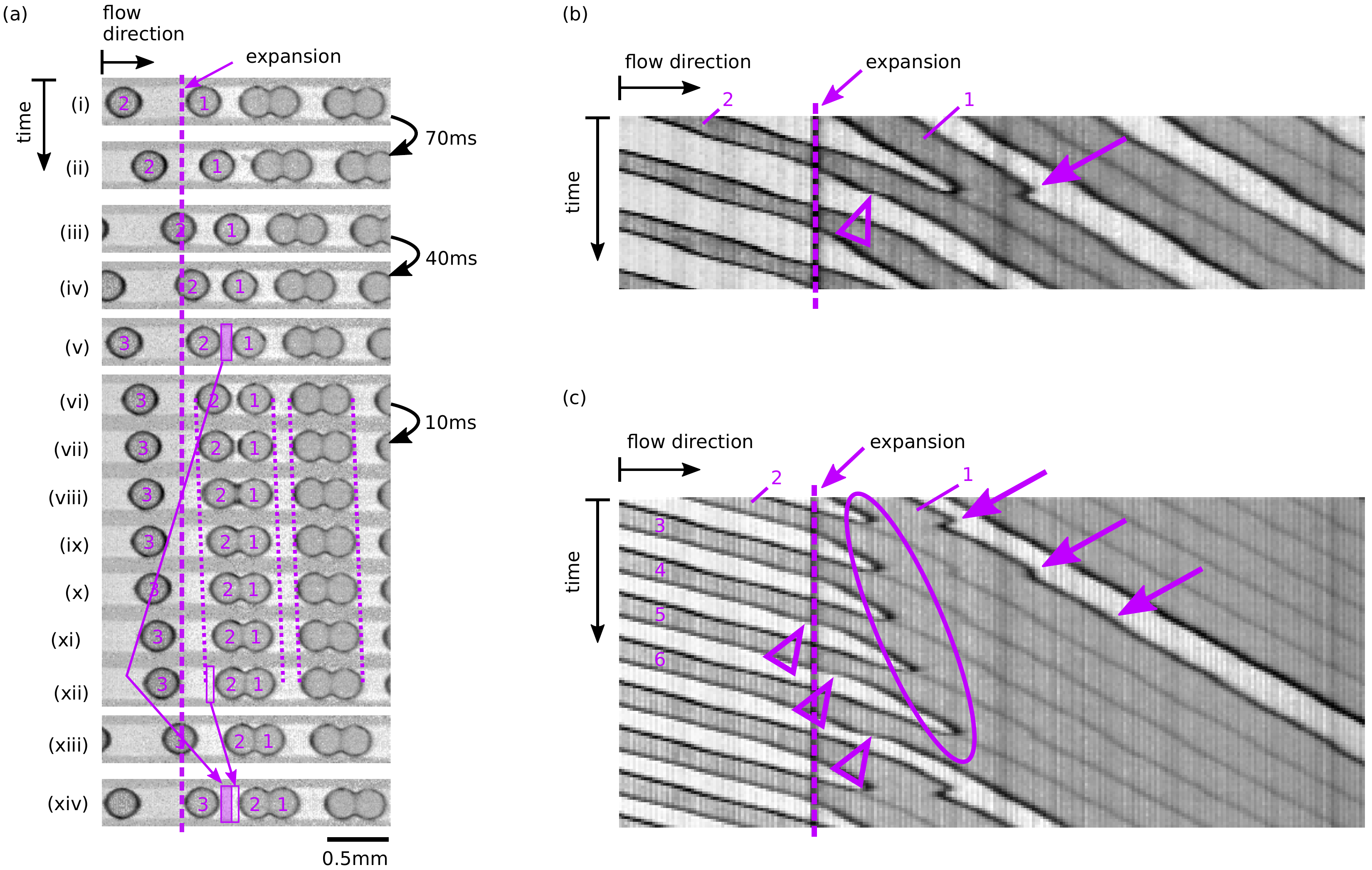}
	\caption{{(a) Time-sequence of images detailing the formation of a cluster of two coated microdroplets (labelled 1 and 2) downstream of the expansion (vertical dashed line). The time interval between images is varied between \SI{70}{\milli\second} and \SI{10}{\milli\second} to highlight key events. The filled purple rectangle indicates the separation distance between droplet~1 and 2 at the instant when droplet~2 clears the expansion and is therefore moving with a similar velocity as droplet~1. The dotted lines show the displacement of the two-droplet system due to the flow of the outer phase, with the slope of these lines indicating a constant translation velocity. The empty purple rectangle shows the additional displacement of droplet~2 towards droplet~1 following the coalescence of their coating films. Once the next droplet labelled~3 clears the expansion, the separation between droplet 3 and the two-droplet cluster (formed of droplets~1 and~2) is indicated by the combined widths of the filled and empty purple rectangles. (b, c) Spatio-temporal diagrams showing the formation of clusters of two and six droplets, respectively. The dark and light grey bands indicate the droplet's inner and the outer phase, respectively. The dark lines correspond to the coating films. The numbers in the figures refer to the order in which the microdroplets sequentially join the cluster. The arrows point to kinks in the front of the cluster front indicating the relative backward motion driven by coalescence. The triangles point to distortions in the bands indicating the excess forward motion of the joining droplet during coalescence.}}\label{fig:2process+xts}
\end{figure*}

Figure~\ref{fig:2process+xts}(a) details the formation of two-droplet clusters with a series of snapshots separated by varying time intervals. In (a), droplet~1 has cleared the expansion and thus, its velocity has reduced to its value in the expanded channel because inertial effects are insignificant. Meanwhile, droplet~2 is advancing at a larger velocity in the upstream channel in (i--iv) thereby reducing its distance to droplet~1. When droplet~2 clears the expansion in (v) and thus adopts a similar velocity to droplet 1, the downstream separation distance between the two droplets $d_{\rm d}$, indicated by the width of a filled purple rectangle, is sufficiently small for the two droplets to react on the attractive forces within that region that arise from hydrodynamic interaction in the microfluidic droplet train \citep{Beatus2006}. While the two-droplet system propagates steadily due to the flow of the outer phase, this attractive force results in the elongation of the droplets (vi, vii) until their coating films coalesce to form a bridge (vii, viii). This is followed by a surface-tension-driven reorganisation of the coating films, which promotes the rapid displacement of the droplets towards each other (viii--xii) until a two-droplet cluster configuration is reached (xii). In the meantime, droplet~3 is catching up with the preceding group by propagating faster in the upstream channel (v--xiii). Once it clears the expansion (xiv), its separation from the rear of the cluster corresponds to the combined widths of the filled purple rectangle that separated droplets~1 and 2 at a similar stage and the empty purple rectangle that indicates the forward displacement of droplet~2 in response to the capillary forces driving the cluster formation. This distance is large enough so that the two-droplet cluster and droplet~3 are not attracted towards each other and continue to propagate downstream with constant separation, while droplet~3 becomes the first droplet of the next cluster. 

The two-droplet cluster formation is summarised in a spatio-temporal diagram in figure~\ref{fig:2process+xts}(b). The dark and light bands indicate the droplets of inner-phase and the outer-phase, respectively, while the dark lines correspond to the coating films. As discussed in section~\ref{sec:visualisation}, the straight bands of approximately constant width upstream of the expansion indicate a regular, single-droplet train. The widening of the dark band immediately downstream of the expansion indicates the reduction in velocity as the droplet passes over the expansion. When droplet 2 catches up with droplet 1 so that their separation is smaller than a critical distance $d_{\rm crit}$, their coating films coalesce. This is indicated by the two dark bands merging so that they are only separated by a thin darker line of coating. Following the coalescence of the coating films, the kink highlighted by an arrow indicates the backward motion of droplet 1, while the triangle points to a distortion of the dark band of droplet 2 which signals a rapid forward movement. The two-droplet cluster then rapidly settles into its final configuration and propagates with constant velocity. 

The spatio-temporal diagram in figure~\ref{fig:2process+xts}(c) details the formation of a six-droplet cluster. The sequential coalescence with the the rear of the cluster takes place as soon as each droplet approaches the cluster to within $d_{\rm crit}$. For the first three droplets (2, 3 and 4), this happens before each of the droplets have cleared the expansion. In fact, the location of the coalescence event is displaced further downstream from the expansion as the cluster grows (see purple ellipse). This is because each time a droplet is added to the cluster, the next droplet needs to travel further to catch up with the rear of the cluster because of previous surface-tension-driven reconfiguration of droplets within the forming cluster. For each successive droplet joining the cluster, the decreasing kinks pointed to by arrows indicate that the backward motion of the cluster is suppressed and in fact, it is not measurable beyond the fourth droplet. In contrast, the forward displacement of the joining droplet is enhanced particularly from the fourth droplet as highlighted by the pointing triangles. This is because the viscous resistance to the displacement of a cluster scales approximately with the length of the cluster, i.e. the number of constitutive droplets. Hence, the displacements of the joining droplet and cluster as they move towards each other are approximately inversely proportional to the number of droplets that make them up.

We found an approximately constant value of the distance at which droplets or clusters coalesced, $d_\mathrm{crit}=\SI{82}{\micro\meter}\pm \SI{10}{\micro\meter}$, across most experiments and channels. Exceptions were for $\alpha = 1.5$, where droplets with very thin coating co-existed at smaller distances and droplets with the thickest coating coalesced at larger distances. For all other droplets coalescence always occurred if the distance between the front of the joining droplet and the rear of any cluster reached $\SI{82}{\micro\meter}\pm \SI{10}{\micro\meter}$ by the time the joining droplet has cleared the expansion.

\begin{figure*}
\includegraphics[width=\textwidth]{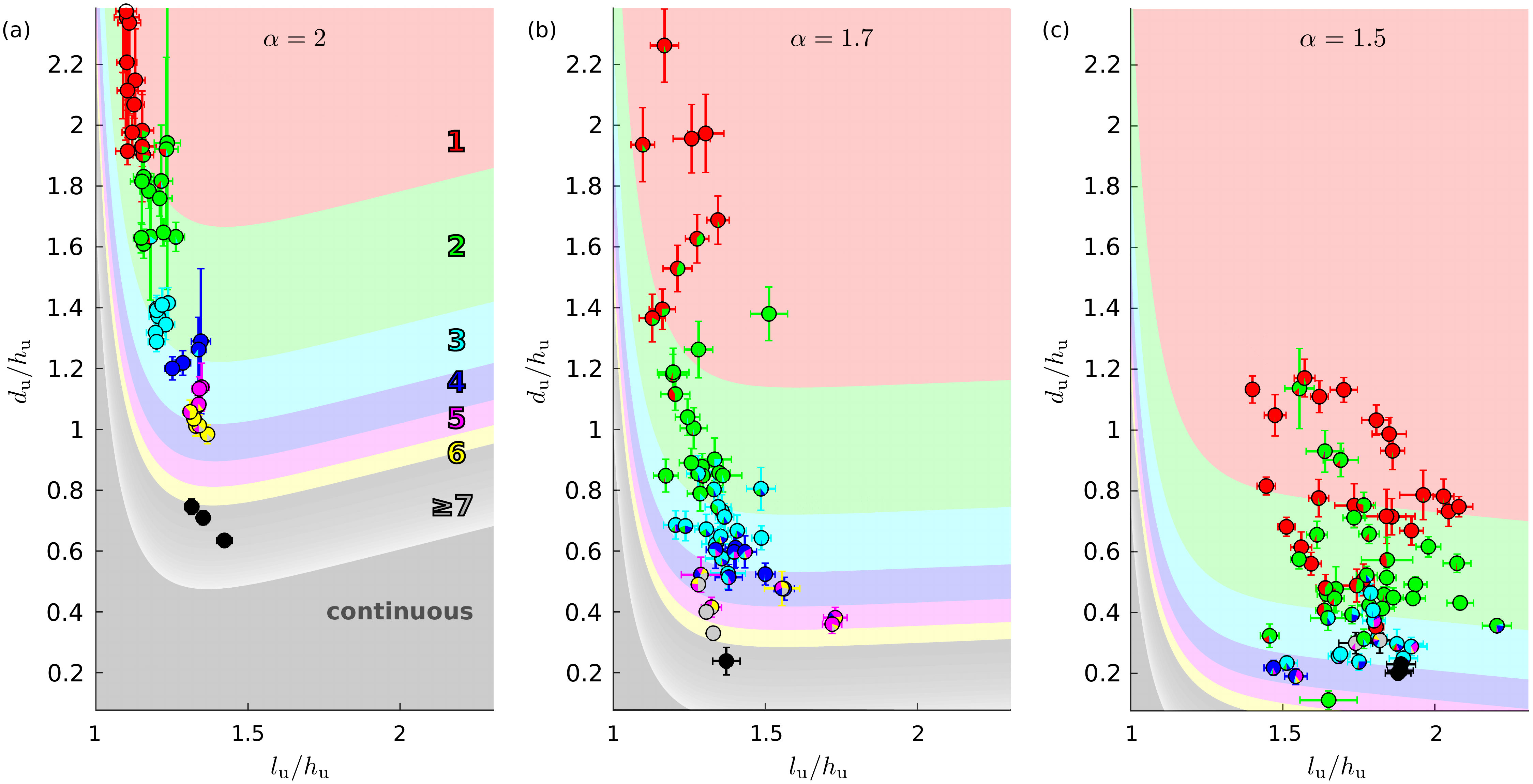}

	\caption{{Phase diagrams of the cluster size downstream of the channel expansion in the parameter plane spanned by inter-drop distance and droplet length (upstream of the expansion). (a) $\alpha=2.0$; (b) $\alpha=1.7$; (c) $\alpha=1.5$.
Experimental data points (symbols) are pie charts showing the fractions of different cluster sizes within the experiment. Each colour corresponds to a different cluster size indicated by the numbers in (a). The error bars represent the standard deviations of droplet length and inter-drop distance within each experiment. The coloured bands indicate the predictions of the toy model discussed in section~\ref{sec:model}. }}\label{fig:comp}
\end{figure*}

The results from 184 experiments performed in micro-channels with $\alpha= \num{2.0}$, \num{1.7} and \num{1.5} are summarised in figure \ref{fig:comp}(a--c), with phase diagrams showing different cluster sizes as a function of inter-drop distance and droplet length. The number of droplets per cluster is indicated by coloured numbers in (a). The coloured bands correspond to predictions of the toy model discussed in section~\ref{sec:model}. 

The variation of inter-drop distance appears to dominate the selection of the cluster size, although this is also the experimental parameter for which we can access the largest span as discussed in section~\ref{sec:parameter}. Overall, the number of droplets per cluster increases as the inter-droplet distance decreases. For $\alpha =2.0$ (a) and $\alpha=1.7$ (b), experimental results are grouped by cluster size with each colour occupying distinct simply-connected regions of the parameter plane.  For $\alpha=1.5$ (c), the regions occupied by red and green symbols overlap but the trend remains similar overall. Outliers (e.g., red circles in the green region and green circles in the red region) correspond to droplets with either very thin or thick coating, for which the value of $d_{\rm crit}$ is likely to differ. The multicoloured symbols correspond to individual experiments where multiple cluster sizes occurred -- the pie chart colouring of the circle indicates the fraction of experiments yielding each cluster size. These mostly occur at the boundaries between regions corresponding to different cluster sizes. 
  
  Decreasing the expansion factor compresses the phase diagrams to span a smaller range of inter-drop distances. For example, for $d_\mathrm{u}/h_\mathrm{u}=\num{1.2}$, clusters of 4 were obtained for $\alpha=2$, clusters of 2 for $\alpha=1.7$ and for $\alpha=1.5$ a single-droplet train was retained downstream of the expansion. This is because the smaller expansions are associated with a smaller drop in velocity downstream of the expansion, thus requiring more closely spaced droplets upstream of the expansion for a droplet to catch up with a cluster to within $d_{\rm crit}$. This compression of the phase diagram was accompanied by a reduction in the maximum cluster size that could be formed in a regular train downstream of the expansion when $\alpha$ was reduced, e.g., four droplets for $\alpha=1.5$. We will revisit these results in section~\ref{sec:model}

\begin{figure}
	\centering
	\includegraphics[width=\columnwidth]{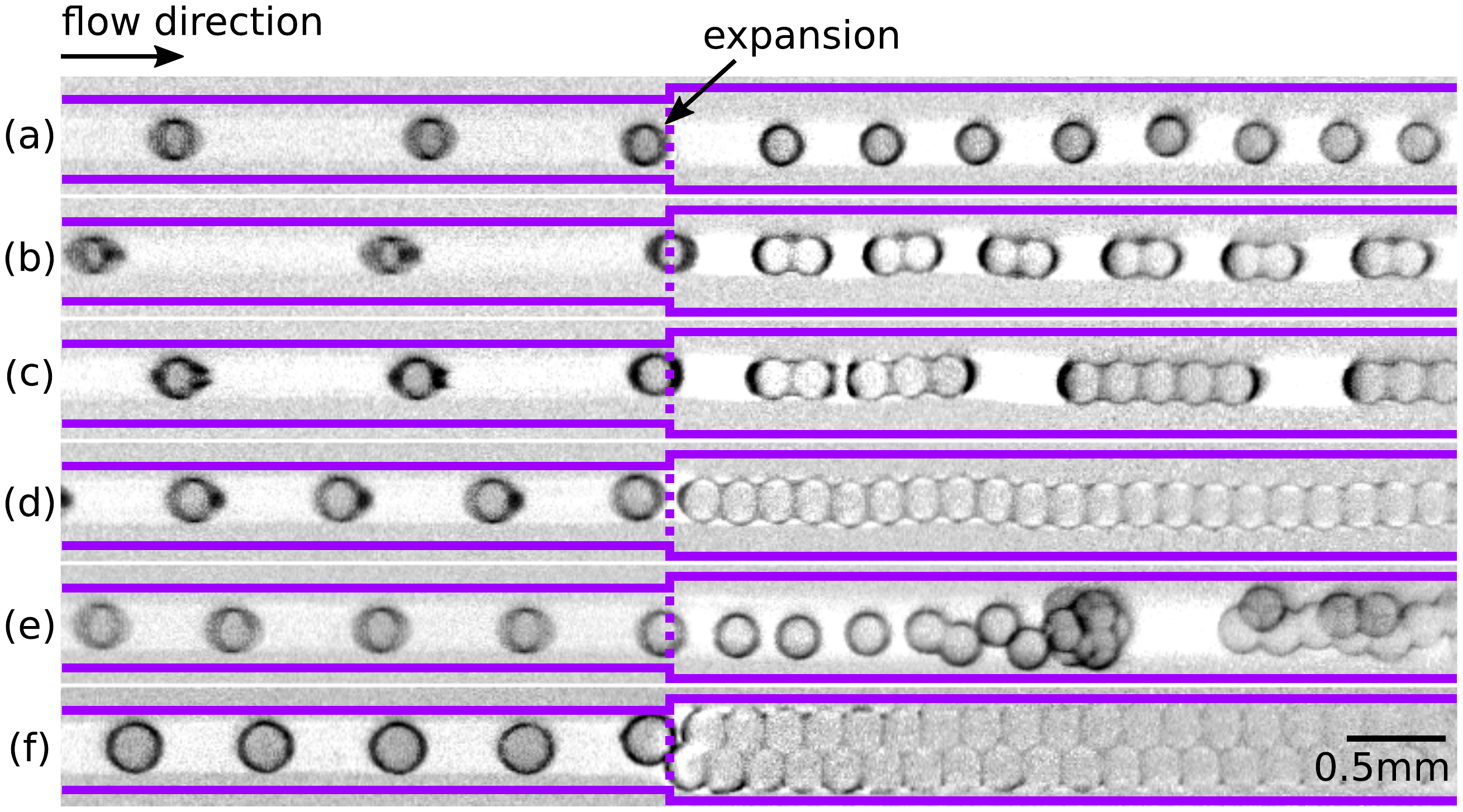}
	\caption{{Series of snapshots illustrating the behaviour of trains of coated microdroplets flowing through a microchannel with a sudden expansion of factor $\alpha=3.0$, indicated by a vertical dotted line. As in figure \ref{fig:gall}, the single-droplet train upstream of the expansion is regular but the train parameters -- inter-drop distance, droplet length, coating film thickness and droplet velocity -- vary between images. The most significant variation is in the inter-drop distance which varies non-monotonically from (a) to (f). The channel boundaries are indicated with purple lines. Background subtraction and contrast adjustment have been applied to all images resulting in some shade variations of the outer phase.}} \label{fig:lambda3}
\end{figure}

Increasing the expansion factor to $\alpha=3$ led to a wider range of droplet dynamics, as shown in figure \ref{fig:lambda3} where all the droplet-train parameters vary between images as in figure \ref{fig:gall}. Although the most significant variation is in the inter-drop distance, this parameter does not decrease monotonically from (a) to (f). For example, the single droplet train retained downstream of the expansion  in (a) occurs for a smaller inter-drop distance than the regular clusters of two droplets in (b). In both cases, these single files of droplets were prone to local distortion with droplets deviating from their path along the channel centreline in a manner similar to the waves excited in one-dimensional microfluidic crystals by stationary defects \citep{Beatus2012}.
In (c), clusters of five droplets are generated in two stages, with the formation of alternate three-droplet and two-droplet clusters and their subsequent coalescence into clusters of five droplets as they propagate downstream. These were the longest finite-size clusters observed in a regular train and a further reduction in the inter-drop distance led to a continuous cluster of droplets connected through their coating film as shown in (d). In contrast with the other expansion factors investigated, $\alpha=3.0$ meant that the cross-sectional dimensions of the downstream channel were sufficient to accommodate at least two typical droplets side-by-side. In (e), the train of droplets forming downstream of the expansion appears disordered with evidence of three-dimensional droplet assemblies. By contrast, in (f) the expansion enables self-assembly of droplets into a two-droplet wide continuous cluster, where droplets of the inner phase are separated by thin coating films. These are reminiscent of ordered emulsions~\citep{Seo2005} and foams~\citep{Marmottant2009}. Hence, the droplet dynamics in the microchannel with $\alpha=3.0$ offer interesting prospects for three-dimensional self-assembly. 

\subsection{Interpretation}
\label{sec:model}

\begin{figure*}
\includegraphics[width=\textwidth]{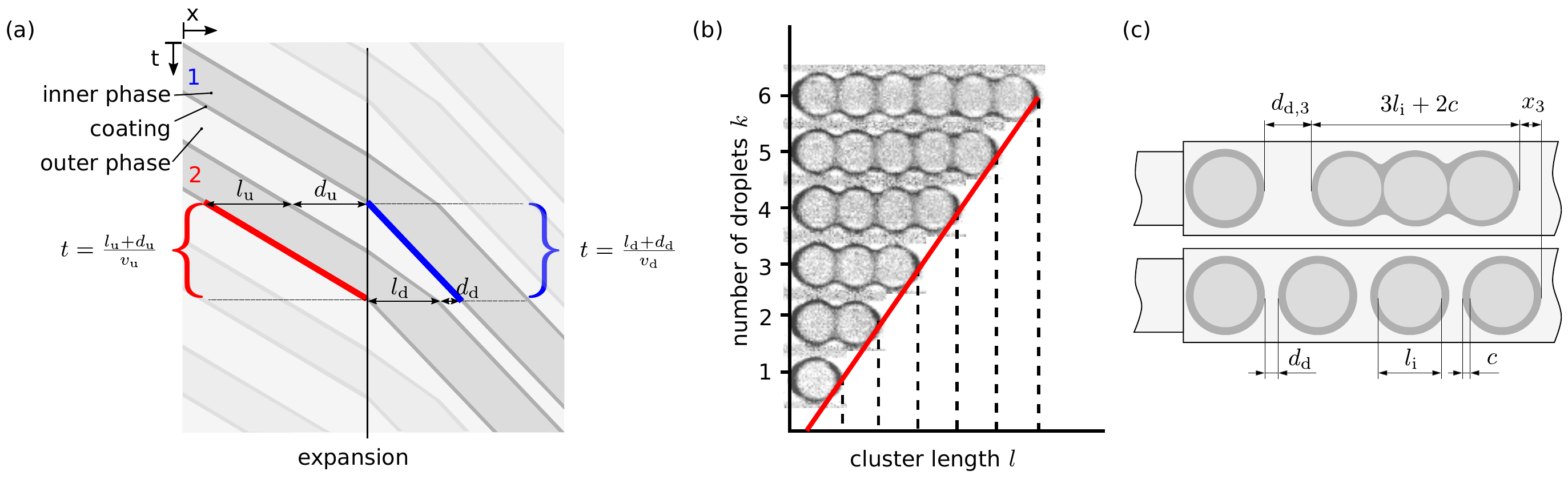}
\caption{{(a) Schematic spatio-temporal diagram near the sudden channel illustrating how the downstream inter-droplet distance $d_{\mathrm{d}}$ is calculated. (b) Sequential formation of a six-drop cluster where the cluster length $l$ is shown as a function of the number of constitutive droplets. The cluster length increases linearly with each added droplet and is accurately captured by the relation $l=k l_\mathrm{i}+2c$ (red line) with $k$ the number of droplets, $l_\mathrm{i}$ the inner droplet length and $c$ the coating thickness at the front and rear of the cluster. (c) Comparison between a droplet following a 3-droplet cluster as per experimental observations and the equivalent regular droplet train.}}\label{fig:dd_dpass}
\end{figure*}

We now formalise our experimental observations for the expansion factors $\alpha=1.5-2.0$ in order to explore in further detail how clustering in these channels depends on the properties of the droplet train. Having observed that clustering occurs if droplets approach each other to within a critical distance in the downstream channel, we calculate the inter-drop distance downstream of the expansion. For this, we consider the time elapsed between the first and second droplet clearing the expansion, see the schematic spatio-temporal diagram in figure~\ref{fig:dd_dpass}(a). In that time, the rear of the first droplet travels a distance $l_{\mathrm{d}}+d_{\mathrm{d}}$ with velocity $v_{\mathrm{d}}$ in the downstream channel (blue, right) while the rear of the second droplet travels $l_{\mathrm{u}}+d_{\mathrm{u}}$ with velocity $v_{\mathrm{u}}$ in the upstream channel (red, left). Hence,
\begin{equation}
%d_{\mathrm{d}}=\frac{v_{\mathrm{d}}}{v_{\mathrm{u}}}d_{\mathrm{u}}+(\frac{v_{\mathrm{d}}}{v_{\mathrm{u}}}-\frac{l_{\mathrm{d}}}{l_{\mathrm{u}}})l_{\mathrm{u}}. \label{eq:dd}
d_{\mathrm{d}}=\frac{v_{\mathrm{d}}}{v_{\mathrm{u}}}(d_{\mathrm{u}}+l_{\mathrm{u}})-l_{\mathrm{d}}. \label{eq:dd}
\end{equation}
This quantity depends on both upstream and downstream droplet lengths, upstream inter-drop distance, as well as the ratio of downstream to upstream velocities. 

\begin{figure*}
\includegraphics[width=\textwidth]{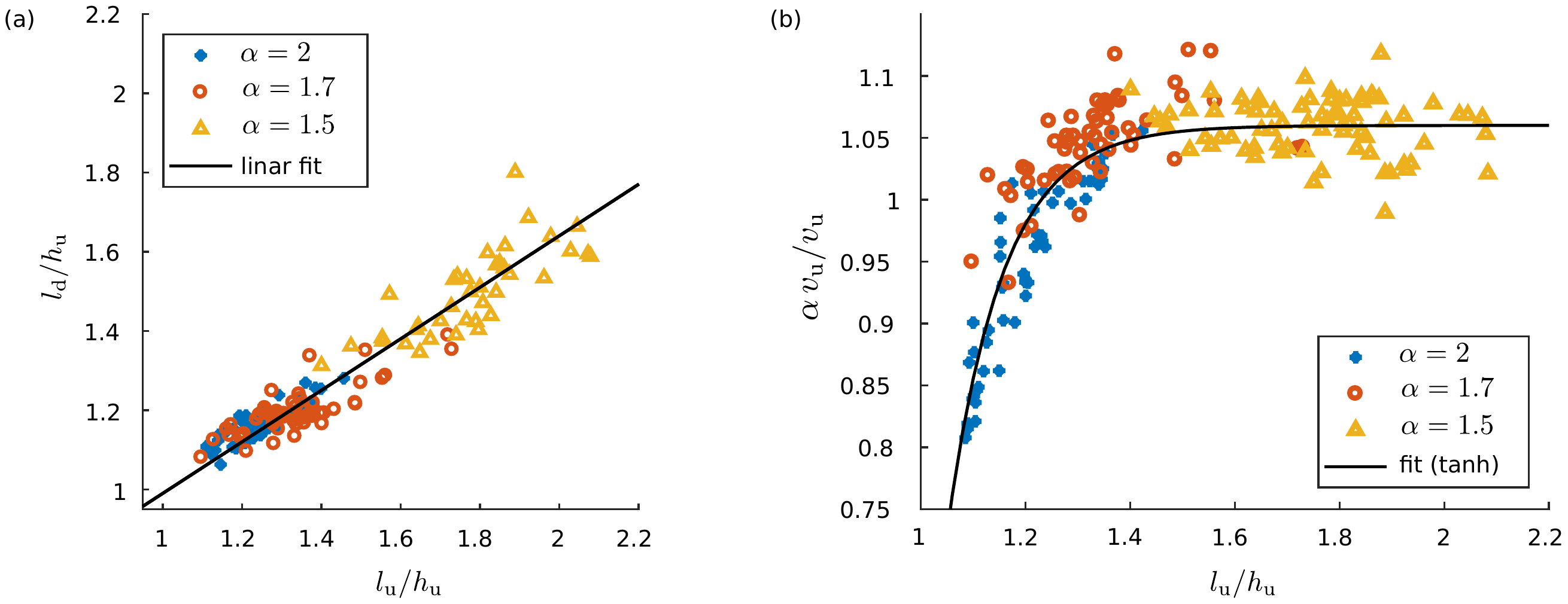}
\caption{{Variation of (a) the scaled downstream droplet length and (b) the ratio of downstream to upstream droplet velocity multiplied by the expansion factor as functions of the scaled upstream droplet length in channels with different expansion factors $\alpha$. Lines represent the fits to the experimental data used in the toy model. (a)~
$l_{\mathrm d}/h_{\mathrm u}= a_1\, l_{\mathrm u}/h_{\mathrm u} +a_0$,
where $a_1=\num{0.65}$ and $a_0=\num{0.35}$; (b)~$\alpha\, v_{\mathrm{d}}/v_{\mathrm{u}} =b_1 (\tanh(b_2 l_{\mathrm{u}}/h_{\mathrm{u}}) -1)+b_3$
with $b_1=3380$, $b_2=4.73$ and $b_3=1.06$}} \label{fig:ratios}
\end{figure*}

 Figure \ref{fig:ratios}(a) shows experimental measurements of the scaled downstream droplet length $l_{\mathrm d}/h_{\mathrm u}$ as a function its upstream value $l_{\mathrm u}/h_{\mathrm u}$. The expansion of the channel allows confined droplets to reduce their length. Whereas we would expect $l_{\mathrm{d}}\simeq l_{\mathrm{u}}$ for droplets that are sufficiently small so that they are unconfined upstream of the expansion, predictions for the intermediate droplet sizes produced in our experiment require direct numerical simulations~\citep{Wang2011}. 
Although for sufficiently long bubbles we would expect the reduction in bubble length to scale approximately with $\alpha$, so that $l_{\mathrm{d}} \simeq l_{\mathrm{u}}/\alpha$, the data shown in figure \ref{fig:ratios}(a) for our intermediate droplet size range is approximately similar for all three values of $\alpha$. Hence, in our toy model, we approximate our data by a least-square linear fit, see figure \ref{fig:ratios}(a).

Figure \ref{fig:ratios}(b) shows that the velocity ratio $v_{\mathrm{d}}/v_{\mathrm{u}}$ rescaled by the expansion factor (or droplet-mobility ratio) collapses  approximately onto a master curve as a function of the upstream droplet length. Droplet mobility is set by the multiphase flow \citep{Sessoms2009, Jakiela2011} and for trains of droplets it also depends on the interaction between droplets~\citep{Wang2011, Jakiela2016}. We find that the mobility ratio increases with droplet size and reaches an approximately constant plateau value for $l_{\mathrm u}/h_{\mathrm u} \ge 1.5$. This means that variations in mobility ratio are significant particularly in the channel with expansion factor $\alpha=2$ because of the smaller droplets sizes in this channel. To capture these variations, we approximate the data in figure \ref{fig:ratios}(b) by a hyperbolic tangent function in our toy model.  

During the sequential formation of a droplet cluster, the surface-tension-driven reconfiguration of the coating films upon addition of a droplet led to an encapsulating film of measurable thickness as well as very thin but apparently stable films separating individual droplets within the cluster. Hence, the length of the growing cluster shown in figure~\ref{fig:dd_dpass}(b) increased linearly as a function of the number of constitutive droplets, so that it was accurately captured by the relation 
\begin{equation}
l=k l_i+2c,  \label{eq:lk}
\end{equation}
where  $k$ is the number of droplets, $l_i$ the length of a droplet of inner phase and $c$ is the thickness of the encapsulating film which did not vary within measurement resolution.

The transition from single droplets to clusters of two droplets downstream of the expansion occurs when the downstream inter-drop distance $d_{\mathrm{d}}$ is reduced to the threshold value $d_{\mathrm{crit}}$ (see table~\ref{tab:modelvalues}). For larger clusters ($k > 2$), the distance $d_{\mathrm{d,k}}$ from the front of a droplet to the rear of the preceding cluster of $k$ droplets is different from $d_{\mathrm{d}}$ because of the droplet reconfiguration that is associated with each droplet addition to the cluster through coalescence of the coating film. Based on the comparison shown in figure~\ref{fig:dd_dpass}(c) between a droplet following a cluster of three droplets and the equivalent regular droplet train, we calculate the distance between the joining droplet and the cluster of $k$ droplets ($k\ge 2$) to be
\begin{equation}
d_{\mathrm{d},k}= k\: d_{\mathrm{d}} + 2(k-1)c - x_k, \label{eq:ddk_simple}
\end{equation}  
where $x_k$ is the total upstream displacement of a cluster of $k$ droplets during formation prior to droplet $k+1$ joining the cluster.

For a cluster of $k$ droplets ($k\ge 2$) to coalesce with the next droplet, the distance separating the front of the inner joining droplet from the rear of the cluster is $d_{\mathrm{crit}}+c$. The cluster is displaced by an approximate fraction $1/k$ of this distance because its length is approximately $k$-times that of a single drop and viscous resistance to the motion scales with the length of the cluster. %, e.g. $(d_{\mathrm{crit}}+2c)(1/2+1/3)$ for a cluster of three.
Thus, we have
\begin{equation}
x_k=(d_{\mathrm{crit}}+c)\sum_{i=2}^{k}\frac{1}{i}, \label{eq:xk}
\end{equation}
which yields
\begin{equation}
d_{\mathrm{d},k}= k\: d_{\mathrm{d}}+(2k-1)c+d_{\mathrm{crit}}-(d_{\mathrm{crit}}+c)\sum_{i=1}^{k}\frac{1}{i}
\label{eq:ddk}
\end{equation}

We used equation~(\ref{eq:ddk}) to calculate the number of droplets per cluster by determining the smallest value of $k$ for which the inequality $d_{\mathrm{d},k}>d_{\mathrm{crit}}$ was satisfied. In this case, the next droplet in the sequence would not join the cluster and thus $k$ was the cluster size.
We explored parameter values representative of the experiments by varying the upstream droplet length and the inter-drop distance. Based on our experimental measurements, we imposed constant values for the coating thickness parameter $c$ and $d_{\mathrm{crit}}$ (see table \ref{tab:modelvalues}) while values for $l_{\mathrm{d}}$ and $v_{\mathrm{d}}/v_{\mathrm{u}}$ were provided by the fits to the experimental data shown in figure \ref{fig:ratios}.

\begin{table}
	\caption{Experimental values of the critical distance for coalescence and the coating thickness used in the predictions of figure~\ref{fig:comp}. 
}\label{tab:modelvalues}
	\begin{tabular}{llll}
		\hline\noalign{\smallskip}
		Expansion factor $\alpha$&1.5&1.7&2\\ %\hline
		\noalign{\smallskip}\hline\noalign{\smallskip}
		Critical distance for coalescence $d_{\mathrm{crit}}$ in \si{\micro\meter}&\num{82}&82&82\\ %\hline
		Coating thickness parameter $c$ in \si{\micro\meter}&\num{62}&\num{63}&\num{52}\\ 
\noalign{\smallskip}\hline	
	\end{tabular}
\end{table}

Model predictions compare favourably with the experiments in figure~\ref{fig:comp} where each coloured band indicates the parameter region where the model predicts a specific number of droplets per cluster. 
The model exhibits a similar dependence of the cluster size on the inter-drop distance as in the experiments, with larger clusters formed as the inter-drop distance decreases. However, it does not predict a limit to cluster sizes in contrast with the experiments, where cluster sizes containing more than $6,\; 5$ and $4$ droplets for $\alpha=2.0,\; 1.7$ and $1.5$, respectively, did not form consistently and were typically replaced by continuous clusters. In the model, the bands containing each cluster size narrow with increasing number of droplets per cluster to span intervals of inter-drop distance on the same order as the experimental fluctuations. We hypothesise that this prevents the formation of regular clusters and instead promotes continuous clusters. Furthermore, a reduction in the expansion factor limits the maximum cluster size observed because of the decrease in velocity contrast across the expansion, which means that the formation of large clusters requires smaller inter-drop distances that are more strongly affected by experimental fluctuations or cannot be realised at all without coalescence in the upstream channel. This is the case, for example, of the green data point in figure~\ref{fig:gall}(c) at $l_{\mathrm u}/h_{\mathrm u}=1.65$ and $d_{\mathrm u}/h_{\mathrm u}=0.11$, for which coalescence happened in the inlet channel and which was included in the diagram to illustrate this phenomenon. We ascribe the increasing discrepancy for larger groups at $\alpha=2$ to the sensitivity of the velocity ratio to droplet size variations for small droplets.

\section{Conclusion}
\label{sec:conclusion}
We have used a sudden microchannel expansion to assemble coated microdroplets sequentially into regular clusters of different sizes. The inter-drop distance in the droplet train upstream of the expansion is the most important parameter in  determining the cluster size for a given expansion factor. The formation of finite clusters is enabled by the reconfiguration of the droplet assembly upon coalescence of the coating film, which sequentially increases the distance to the next droplet to clear the expansion. This coalescence-induced separation restricts the length of the coalescence sequence at the expansion, in contrast with closely packed trains of microdroplets, where coalescence-induced separation leads to a cascade of coalescence~\citep{Bremond2008, Jose2012}. The increased upstream to downstream velocity ratios associated with larger expansion factors move the onset of the clustering further upstream of the expansion. Hence, the largest cluster size that can be achieved before all droplets join to form a continuous cluster increases with expansion factor in our experiments.

By formalising our observations in a toy model, we find that all cluster sizes could in principle occur but that the range of inter-drop distances for which clusters of increasing size occur decreases sharply. We hypothesize that the maximum cluster size observed experimentally for each expansion factor is set by the natural experimental fluctuations which prevent longer finite clusters and promote continuous clusters instead.

More complex structures were also produced by increasing the value of $\alpha$ to $3$. However, the regular clusters of coated microdroplets obtained for $\alpha \le 2$ could in turn be used to generate complex three-dimensional colloidal structures \citep{Shen2016} by smoothly expanding the channel while supplying additional outer fluid to keep the structures separated. Moreover, our process of self-assembly of single-core double-emulsion droplets at a sudden expansion provides a simple and flexible alternative to the direct production of double emulsions with several inner droplets using droplet generators, which tend to require very fine control of the flow regime \citep{Utada2005,Nabavi2017,Vladisavljevic2017}.

\section*{Conflict of interest}
The authors declare that they have no conflict of interest.

\bibliographystyle{spbasic}      % basic style, author-year citations

\bibliography{cluster_paper}

\end{document}